\documentclass[9pt,twocolumn]{article}

\usepackage{graphicx}
\usepackage{hyperref}
\usepackage[margin=0.5in]{geometry}

\title{\textbf{Ultrafast single-photon detection based on optical Kerr gates at GHz rates}}

\author{Abdul-Hamid Fattah,$^1$ Assegid Mengistu Flatae,$^1$ Amr Farrag,$^1$ and Mario Agio$^{1,2,*}$\\
1 Laboratory of Nano-Optics and C$\mu$, University of Siegen, 57072 Siegen, Germany\\
2 National Institute of Optics (INO), National Research Council (CNR), 50125 Florence, Italy}

\begin{document}

\maketitle

\begin{abstract}
\textbf{The ultrafast detection of single photons is currently restricted by the limited time resolution (a few picoseconds) of the available single-photon detectors. Optical gates offer a faster time resolution, but so far they have been mostly applied to ensembles of emitters. Here, we demonstrate through a semi-analytical model that the ultrafast time-resolved detection of single quantum emitters can be possible using an optical-Kerr-shutter (OKS) at GHz rates under focused illumination. This technique provides sub-picosecond time resolution, while keeping a gate efficiency at around 85 \%. These findings lay the ground for future experimental investigations on the ultrafast dynamics of single quantum emitters, with implications for quantum nanophotonics and molecular physics.\\}
\end{abstract}

\href{mailto:mario.agio@uni-siegen.de}{* mario.agio@uni-siegen.de\\} 


The generation and the efficient detection of ultrafast single photons is at the forefront of modern optical science and it offers an unprecedented set of capabilities, for instance in the area of quantum information science~\cite{arakawa2020progress,migdall2013single,mosley2008heralded}, high-sensitivity detection~\cite{zhang2019potential,shcheslavskiy2016ultrafast} and remote sensing~\cite{polley2015ultrafast,villa2012spad}. In addition, it provides a powerful analysis tool in the life sciences at the single-molecule level~\cite{wahl2014time,ohnuki2006development,brinks2014ultrafast}.
Implementing time-resolved spectroscopy based on time-correlated single-photon counting (TCSPC) and single-photon detectors is currently limited to a few-picoseconds time resolution~\cite{korzh2020demonstration}. 
Likewise, recent work on the ultrafast detection of photons using a streak camera reports a few-picoseconds time resolution~\cite{wiersig2009direct, assmann2010ultrafast, assmann2010measuring}. 

Faster emission processes cannot be resolved in time using these approaches, e.g., fluorescence in porphyrins~\cite{venkatesh2016ultrafast}, $\beta$-carotene~\cite{kandori1994direct}. Moreover, continuous progress in hybrid quantum systems, i.e., quantum emitters coupled with nanophotonics structures, indicates that sub-picosecond emission timescales are within reach~\cite{chen2012metallodielectric,hoang2016ultrafast,flatae2019plasmon,bogdanov2020ultrafast}. These recent developments demand for novel ultrafast approaches for the investigation of the photodynamics at the single-emitter level.

The combination of ultrafast lasers with nonlinear optical sampling techniques allow to achieve faster time resolutions. For instance, the second-order nonlinear effect that leads to sum (up-conversion) or difference (down-conversion) frequency generation of light in a nonlinear crystal (e.g. KH$_2$PO$_4$, LiNbO$_3$) has been used for this purpose~\cite{zhang2006femtosecond,jimenez1994femtosecond,barth2007infrared,mao2015multi,sajadi2013femtosecond}.
Since the second-order nonlinear effect requires phase matching, it efficiently works for a relatively narrow bandwidth determined by the phase-matching angle of the nonlinear optical crystal. Nevertheless, this narrow-band method can reach time resolutions up to 33 fs~\cite{Kim:08}. 

Nonlinear optical sampling based on the optical Kerr shutter (OKS) offers the desirable time resolution (comparable with the up-conversion technique) and more flexibility in terms of bandwidth and spectrum for most applications. For example, Tan et al. report a time resolution of 85 fs using bismuth borosilicate (30 fs with 800 nm pulses at a repetition rate of 1 kHz)~\cite{tan2008control} and Schmidt et al. report a time resolution of about 100 fs by recording the spectro-temporal behaviour of a white-light continuum and of the fluorescence emission of $\beta$-carotene and an azobenzene derivative~\cite{schmidt2003broadband}. 

In this work we develop a semi-analytical model to investigate the potential of the OKS for the detection of single quantum emitters, which requires a high gating efficiency combined with a high repetition rate. These are the two conditions that so far have hindered the application of an OKS in this context, because the third-order nonlinear effect requires larger pulse energies, which are commonly achieved by amplified systems operating at low repetition rates. Here, we show that a single-photon gating efficiency of around 85 \% at 1 GHz can be obtained using a 3-mm thick bismuth borosilicate (Bi$_2$O$_3$-B$_2$O$_3$-SiO$_2$) as a Kerr medium with a time resolution of about 92 fs (FWHM). We consider an average power of 1.8 W at a  wavelength $\lambda$ = 800 nm, a repetition rate of 1 GHz, and a pulse duration of 80 fs, to exemplify experimental conditions achievable with modern laser systems, and a probe beam also at $\lambda$ = 800 nm to simplify the analysis. Nonetheless, our model is general and applicable to other sets of parameters (see Tab.~\ref{kerrtab}).


The OKS technique is based on the third-order nonlinear optical Kerr effect, which is a change in the refractive index of a material due to an applied electric field. The nonlinear material becomes birefringent, when placed in a strong linearly-polarized light field. Hence, the polarization of light traversing the material is partially rotated~\cite{kinoshita2000efficient}. In practice, a probe signal is propagated through the OKS consisting of a Kerr medium between two crossed polarizers. When the gate pulse is absent, the medium is isotropic (there is no birefringence) and the signal transmission is blocked by the second polarizer. When the gate pulse arrives, the polarization direction of the signal is rotated due to the optical Kerr effect, and some of the light is transmitted through the second polarizer and detected by a sensitive photo detector. The detected signal arises only from the part of the signal overlapping in time with the gate pulse. In this detection scheme, unlike up-conversion, the frequency of the probe signal is unaffected.

\begin{figure}
		\centering
		\includegraphics[width=0.5\textwidth]{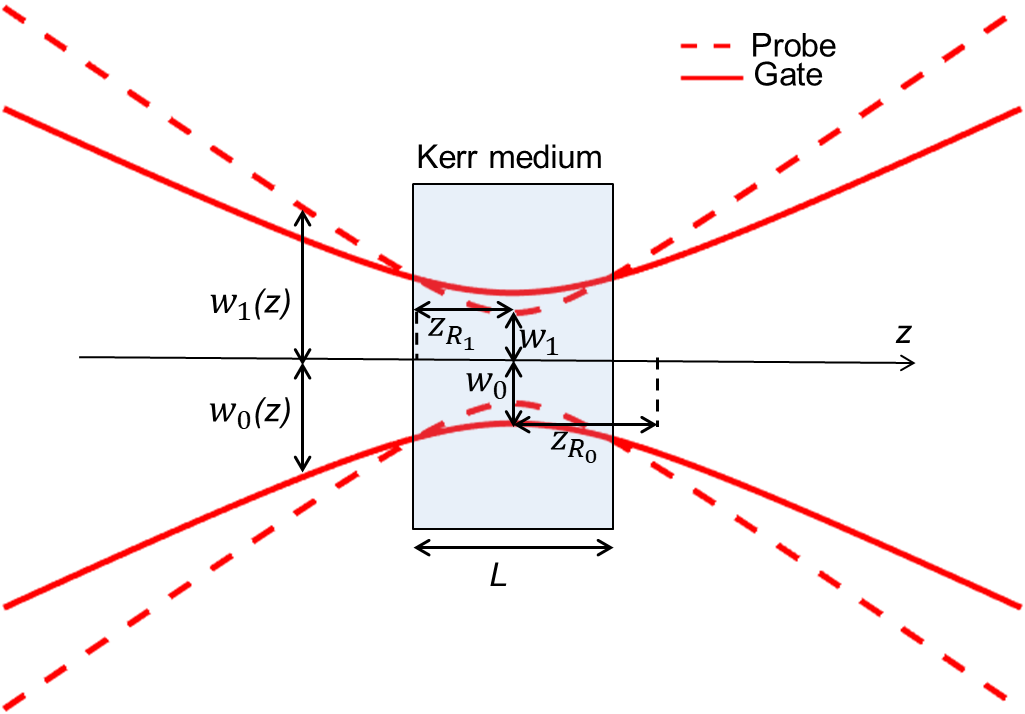}
	\caption{Intensity profiles of the pump (gate) and probe Gaussian beams (without dispersion) in collinear geometry. Both beams are focused inside a Kerr medium of thickness $L$, with different focusing strengths. $\mathrm{w}_0$, $z_{\mathrm{R},0}$ and $\mathrm{w}_1$, $z_{\mathrm{R},1}$ are the beam waist and the Rayleigh range of the gate and probe beams, respectively.}
	\label{OKS}
\end{figure}

 Taking dispersion into account, the intensity profile of the gate pulse in a nonlinear crystal with thickness $L$, as depicted in Fig.~\ref{OKS}, is given in the cylindrical coordinates $r$, $z$ by~\cite{saleh2019fundamentals}
 \begin{equation} \label{1}
 I_\mathrm{g}(r,z)=I_\mathrm{g}(z)\exp{\left[-2\left(\frac{r_0}{w_0(z)}\right)^2\right]},
 \end{equation}
where $\mathrm{w}_0(z)=\mathrm{w}_0\sqrt{1+\left(z/z_{\mathrm{R},0}\right)^2}$, $z_{\mathrm{R},0}=\pi n \mathrm{w}_0^2/\lambda$ is the Rayleigh range, $\mathrm{w}_0$ is the beam waist and $n$ is the linear index of refraction. Figure~\ref{OKS} also shows the Gaussian profile of the probe beam, indicated by the parameters $\mathrm{w}_1$ and $z_{\mathrm{R},1}$.
At $r=0$, the intensity profile takes the form
\begin{equation} \label{20}
I_\mathrm{g}(z)=I_0 I_\mathrm{disp}(z)\frac{1}{1+(z/z_{\mathrm{R},0})^2},
\end{equation}
where $I_0=2\sqrt{\ln{2}}E\tau_0/(\pi^{3/2}w_0^2)$ is the pulse peak intensity, $I_\mathrm{disp}(z)=1/\sqrt{\tau_0^4+[4\ln{2}\phi(z+L/2)]^2}$ the intensity evolution due to material dispersion, $E$ is the pulse energy, $\tau_0$ is the pulse duration, and $\phi$ is the group velocity dispersion (GVD) of the nonlinear crystal.

Since the beam intensity depends on $z$, so does the Kerr efficiency. The corresponding phase-shift induced by birefringence in the nonlinear crystal is given by 
\begin{equation} \label{3}
\mathrm{d}\varphi=(2\pi n n_2/\lambda)I_\mathrm{g}(z)\mathrm{d}z,
\end{equation}
where $n_2$ is the nonlinear index of refraction of the Kerr medium and $\lambda$ is the wavelength of the probe pulse in the medium. The infinitesimal phase-shift must be numerically integrated to obtain the phase-shift $\Delta\varphi(0,L)$ on the optical axis accumulated by the probe beam, when it travels through the nonlinear crystal in the presence of the gate pulse. 
The transmission of the OKS on the optical axis is related to the phase-shift through the formula~\cite{doi:10.1366/0003702053085007}
\begin{equation} \label{4}
T(0,L)=\sin^2\left[\frac{\Delta\varphi(0,L)}{2}\right].
\end{equation}

\begin{figure}[htbp!]
	\centering
	\includegraphics[width=0.5\textwidth]{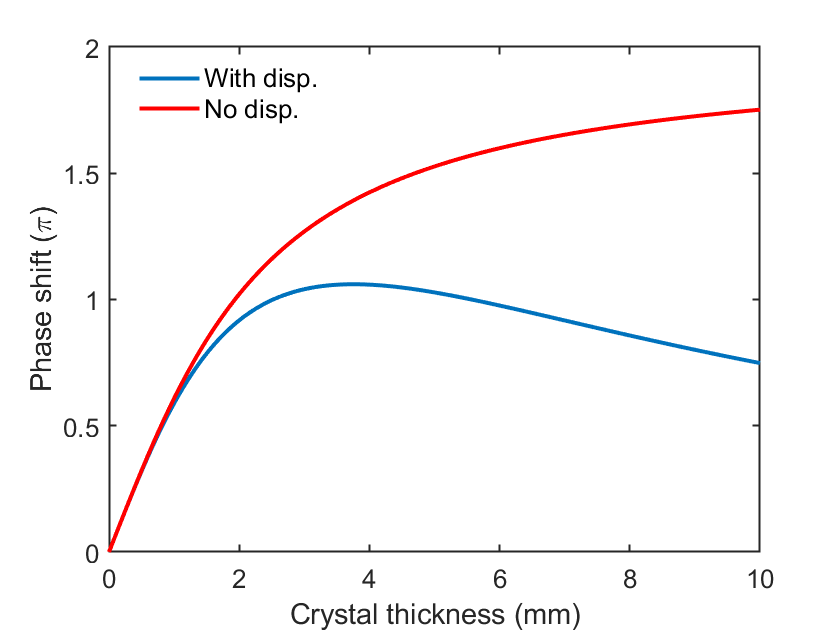}
	\caption{The effect of dispersion on the phase shift $\Delta\varphi(0,L)$ is more apparent for thicker Kerr media or shorter pulses (not shown). Parameters: $\lambda$ = 800 nm, $\mathrm{w}_0$ = 10 $\mu$m, $E$ = 1.8 nJ, $\tau_0$ = 80 fs, $n$ = 2.45, $n_2$ = 1.6$\times10^{-18}$ m$^2$/W, $\phi$ = 1057.19 fs$^2$/mm.}
	\label{disp}
\end{figure}

The effect of dispersion on the phase shift is determined numerically using Eqs.~(\ref{20}) and (\ref{3}). Figure~\ref{disp} shows the phase shift as a function of the crystal thickness with and without dispersion. The effect of dispersion is manifested mainly for thicker crystals of size $L > 2$ mm.

If the pump and probe beams are not collinear, we can calculate the efficiency by defining the effective interaction length of the two pulses inside the medium, which is given by  $L_{\mathrm{eff}}=L\cos(\theta)$, where $\theta$ is the angle between the two beams. 

\begin{figure*}[htbp!]
	\centering
	\includegraphics[width=0.75\textwidth]{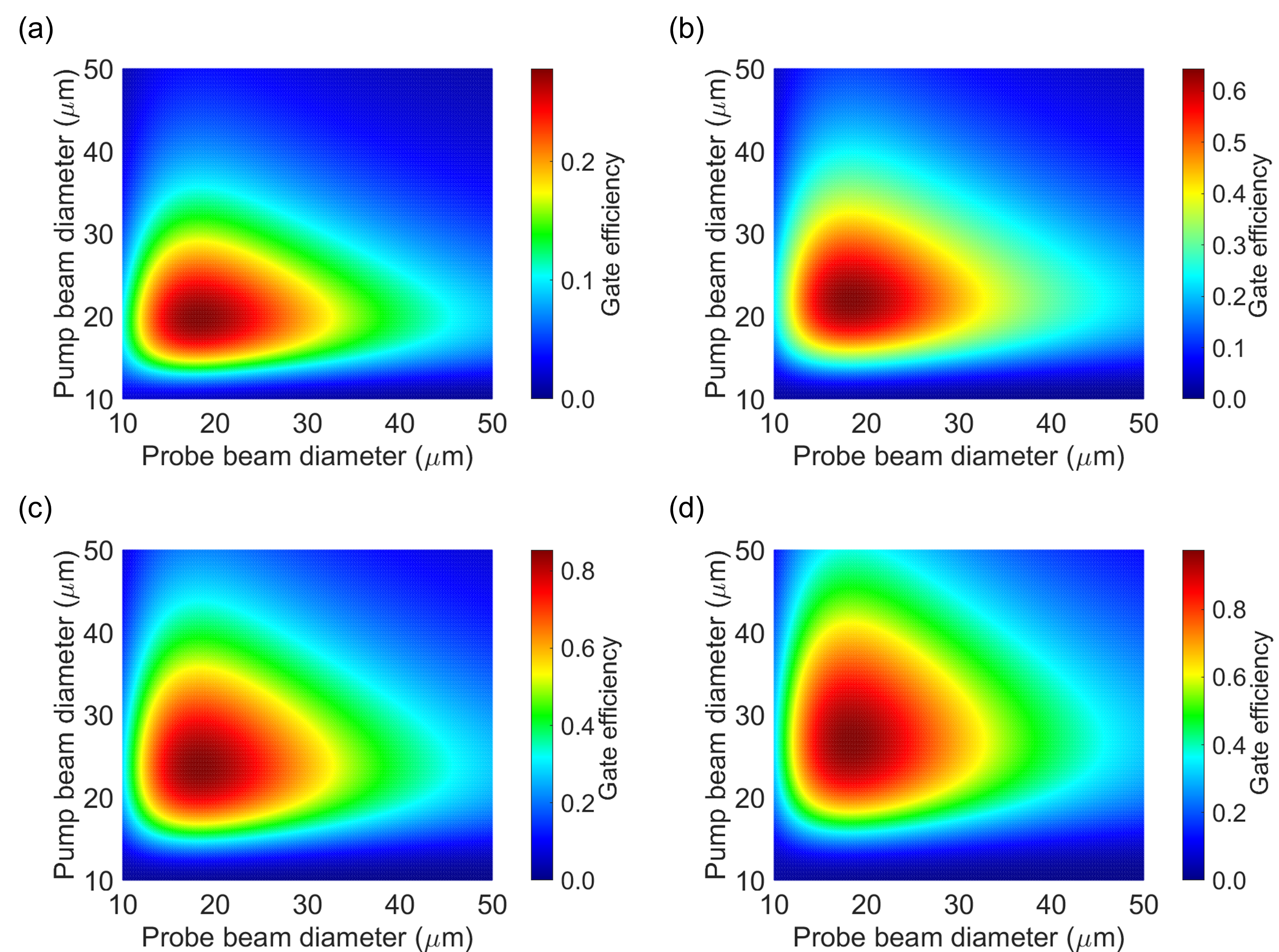}
	\caption{Transmittance (gate efficiency) as a function of the pump and probe beam diameter for different lengths of the nonlinear crystal. (a) $L$ = 1 mm, (b) $L$ = 2 mm, (c) $L$ = 3 mm, (d) $L$ = 5 mm. Parameters: $\lambda$ = 800 nm, $E$ = 1.8 nJ, $\tau_0$ = 80 fs, $n$ = 2.45, $n_2$ = $1.6\times10^{-18}$ m$^2$/W, $\phi$ = 1057.19 fs$^2$/mm.}
	\label{pump-probe}
\end{figure*}

The determination of the OKS efficiency is obtained by taking into account the fact that the photon and gate pulses have a Gaussian profile. In practice, every ray composing the probe beam experiences a different phase-shift as the intensity is a function of not only $z$, but also of the ray trajectory defined by
\begin{equation} \label{5}
r(z)=r_0\sqrt{1+(z/z_\mathrm{R})^2},
\end{equation}
where $r_0$ is the distance from the optical axis at the beam waist (the same expression holds for the pump and probe beams, provided the respective Rayleigh range is used). This leads to a phase-shift defined by 
 \begin{equation} \label{7}
\Delta\varphi(r_0,L)=\Delta\varphi(0,L)\exp{\left[-2\left(\frac{r_0}{w_0}\right)^2\right]},
\end{equation}
where the phase-shift $\Delta\varphi(0,L)$ without dispersion takes the form
\begin{equation} \label{8}
\Delta\varphi(0,L)=\frac{4\pi n n_2 z_{\mathrm{R},0} I_0}{\lambda} \arctan\left(\frac{L}{2z_{\mathrm{R},0}}\right).
\end{equation}

According to the previous argument, the transmittance for the ray $r_0$ is given by 
\begin{equation} \label{9}
T(r_0,L)=\sin^2\left[\frac{\Delta\varphi(r_0,L)}{2}\right].
\end{equation}

Hence, the transmittance of the probe pulse is the average transmittance of all rays defined by
\begin{equation} \label{21}
T(w_0,w_1,L)=\frac{\int_{0}^{\infty}T(r_0,L)\exp({-2r_0^2/w_1^2})r_0 \mathrm{d}r_0}{\int_{0}^{\infty}\exp({-2r_0^2/w_1^2})r_0 \mathrm{d}r_0},
\end{equation}
which is a function of the waist of the pump beam ($w_0$) and of the probe beam ($w_1$). The outcome of this integral is plotted in Fig.~\ref{pump-probe} as a function of different beam diameters and for different crystal thicknesses. The results show that by optimizing the probe and pump beam diameter, the OKS provides more than 85\% single-photon gating efficiency and that there is a correlation between the two waists. Simulations performed under different conditions (not shown) report a similar correlation.

\begin{figure}[htbp!]
	\centering
	\includegraphics[width=0.5\textwidth]{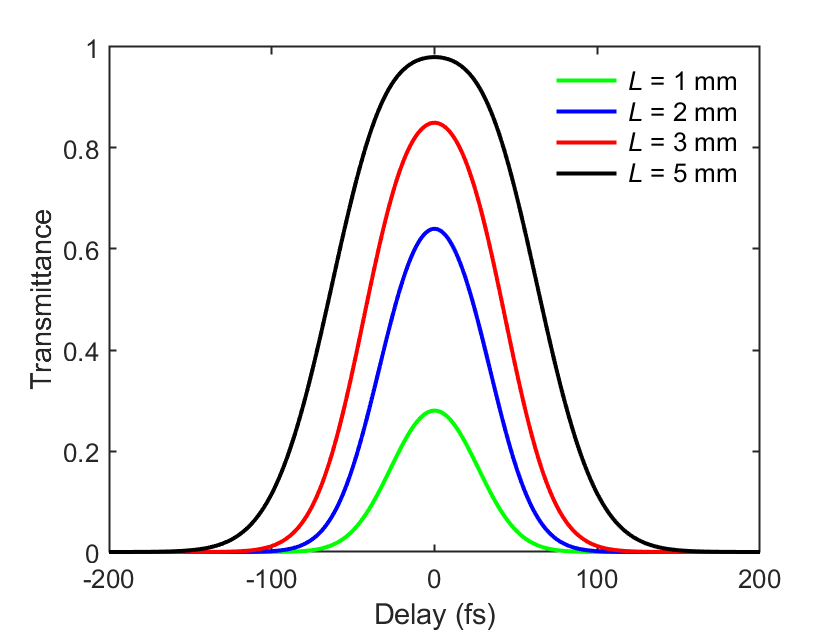}
	\caption{Transmittance as a function of time for different thicknesses of the nonlinear crystal. Parameters: $\lambda$ = 800 nm, $E$ = 1.8 nJ, $\tau_0$ = 80 fs, $n$ = 2.45, $n_2$ = 1.6$\times10^{-18}$ m$^2$/W, $\phi$ = 1057.19 fs$^2$/mm.}
	\label{timeres}
\end{figure}

The time resolution defining the instrument response function is determined by the gate pulse intensity in the time domain according to
\begin{equation} \label{30}
I_\mathrm{g}(t)=I_0\exp\left[-\frac{(t-t_0)^2}{2\tau_0^2}\right].
\end{equation}
Figure~\ref{timeres} depicts the time resolution of the gate for different thicknesses of the nonlinear crystal, which has been obtained using Eq.~(\ref{4}) assuming a constant pump beam intensity that corresponds to the gate efficiency calculated by Eq.~(\ref{21}). In general, the time resolution decreases as the crystal thickness increases. In addition, dispersion plays a role by further reducing the time resolution as it is apparent for thicker crystals. 

\begin{table*}
\centering
\resizebox{0.75\textwidth}{!}{\begin{tabular}{c|c|c|c|c|c|c|c|c|c|c}
\begin{tabular}[c]{@{}c@{}}Kerr \\medium \end{tabular} & \begin{tabular}[c]{@{}c@{}}$n_2$ \\(m$^2$/W) \end{tabular} &
\begin{tabular}[c]{@{}c@{}}$L$ \\ (mm) \end{tabular} & n              & \begin{tabular}[c]{@{}c@{}}$\phi$ \\(fs$^2$/mm)\end{tabular} & \begin{tabular}[c]{@{}c@{}}$2w_0$ \\($\mu$m)\end{tabular} & \begin{tabular}[c]{@{}c@{}}$2w_1$ \\($\mu$m)\end{tabular} & \begin{tabular}[c]{@{}c@{}}$\tau_0$ \\ (fs) \end{tabular} & \begin{tabular}[c]{@{}c@{}}Rep. rate \\ (MHz) \end{tabular} & \begin{tabular}[c]{@{}c@{}}OKS \\ (Eff.) \end{tabular} & \begin{tabular}[c]{@{}c@{}}OKS \\ (FWHM) \end{tabular}  \\ 
\hline
BBS                         & $1.6\times10^{-18}$                                                     & 3                                                     & 2.45  & 1057.19                                                 & 23.7                                                                              & 18.5                                                                                 & 80                                                     & 10$^3$                                                 & 85 \%                                                      & 92 fs                                                                 \\
BBS                         & $1.6\times10^{-18}$                                                     & 0.1                                                   & 2.45  & 1057.19                                                 & 18.5                                                                              & 18.5                                                                                 & 60                                                     & 80                                                & 85 \%                                                      & 51 fs                                                                 \\
SiO$_2$                                           & $2.7\times10^{-20}$                                                     & 2                                                     & 1.47    & 36.16                                                   & 24.1                                                                              & 18.5                                                                                 & 60                                                     & 10                                                & 92 \%                                                      & 54 fs                                                                
\end{tabular}}
\caption{Efficiency and time resolution (FWHM) of the OKS for different Kerr media, such as bismuth borosilicate (Bi$_2$O$_3$-B$_2$O$_3$-SiO$_2$, BBS)~\cite{tan2008control,lightcon} and fused silica (SiO$_2$)~\cite{karras2014nonlinear} and various experimental parameters: $\lambda$ = 800 nm, average pump power 1.8 W, and the other ones as listed in the table. An instantaneous nonlinear material response is assumed. \label{kerrtab}}
\end{table*}

Table~\ref{kerrtab} summarizes the single-photon gating efficiency of Kerr media under different experimental conditions, which has been calculated using our semi-analytical model. It has been estimated that a OKS based on a 3-mm thick bismuth borosilicate exhibits a time resolution of around 92 fs with an efficiency of around 85 \%. Unlike previously reported work, an OKS shutter at GHz repetition rates enables a high single-photon gating efficiency, while maintaining sub-picosecond scale time resolution, provided that the gate and probe beams are correctly focused. 

In conclusion, we  demonstrate through a semi-analytical model that the time-resolved detection of ultrafast single photons is possible using an OKS at GHz rates. This technique provides sub-picosecond time resolution, while keeping a gate efficiency above 85\%, provided that the pulses are focused down to a few tens of microns. Moreover, we show that the focusing strength of the pump and probe beams is correlated and that it is somewhat independent of other parameters (see Tab.~\ref{kerrtab}).
These findings lay the ground for future experimental investigations of ultrafast photophysical processes at the single-emitter level, which has become relevant for recent developments in quantum nanophotonics, and which will enable more insight on fundamental studies in molecular physics.

The authors gratefully acknowledge financial support from the University of Siegen and the German Research Foundation (DFG) (INST 221/ 118-1 FUGG). M. Agio would like to thank P. Foggi and T. Lenzer for helpful discussions.








\begin{thebibliography}{99}

\newcommand{\enquote}[1]{``#1''}

\bibitem{arakawa2020progress}
Y.~Arakawa and M.~J. Holmes, \enquote{Progress in quantum-dot single photon
  sources for quantum information technologies: A broad spectrum overview,}
  {Applied Physics Reviews}, \textbf{7}, 021309 (2020).

\bibitem{migdall2013single}
A.~Migdall, S.~V. Polyakov, J.~Fan, and J.~C. Bienfang, \emph{Single-photon
  generation and detection: physics and applications} (Academic Press, 2013).

\bibitem{mosley2008heralded}
P.~J. Mosley, J.~S. Lundeen, B.~J. Smith, P.~Wasylczyk, A.~B. U’Ren,
  C.~Silberhorn, and I.~A. Walmsley, \enquote{Heralded generation of ultrafast
  single photons in pure quantum states,} {Physical
  Review Letters} \textbf{100}, 133601 (2008).

\bibitem{zhang2019potential}
H.~Zhang, L.~Xiao, B.~Luo, J.~Guo, L.~Zhang, and J.~Xie, \enquote{The potential
  and challenges of time-resolved single-photon detection based on
  current-carrying superconducting nanowires,} {Journal
  of Physics D: Applied Physics} \textbf{53}, 013001 (2019).

\bibitem{shcheslavskiy2016ultrafast}
V.~Shcheslavskiy, P.~Morozov, A.~Divochiy, Y.~Vakhtomin, K.~Smirnov, and
  W.~Becker, \enquote{Ultrafast time measurements by time-correlated single
  photon counting coupled with superconducting single photon detector,}
 {Review of Scientific Instruments} \textbf{87}, 053117
  (2016).

\bibitem{polley2015ultrafast}
N.~Polley, S.~Singh, A.~Giri, P.~K. Mondal, P.~Lemmens, and S.~K. Pal,
  \enquote{Ultrafast fret at fiber tips: Potential applications in sensitive
  remote sensing of molecular interaction,} {Sensors and
  Actuators B: Chemical} \textbf{210}, 381--388 (2015).

\bibitem{villa2012spad}
F.~Villa, D.~Bronzi, S.~Bellisai, G.~Boso, A.~B. Shehata, C.~Scarcella,
  A.~Tosi, F.~Zappa, S.~Tisa, D.~Durini \emph{et~al.}, \enquote{Spad imagers
  for remote sensing at the single-photon level,} in \emph{Electro-Optical
  Remote Sensing, Photonic Technologies, and Applications VI,}  vol. 8542
  (International Society for Optics and Photonics, 2012), p. 85420G.

\bibitem{wahl2014time}
M.~Wahl, \enquote{Time-correlated single photon counting,}
  {Technical Note} pp. 1--14 (2014).

\bibitem{ohnuki2006development}
T.~Ohnuki, X.~Michalet, A.~Tripathi, S.~Weiss, and K.~Arisaka,
  \enquote{Development of an ultrafast single photon counting imager for single
  molecule imaging,} in \emph{Ultrasensitive and Single-Molecule Detection
  Technologies,}  vol. 6092 (International Society for Optics and Photonics,
  2006), p. 60920P.

\bibitem{brinks2014ultrafast}
D.~Brinks, R.~Hildner, E.~M. Van~Dijk, F.~D. Stefani, J.~B. Nieder,
  J.~Hernando, and N.~F. Van~Hulst, \enquote{Ultrafast dynamics of single
  molecules,} {Chemical Society Reviews} \textbf{43},
  2476--2491 (2014).

\bibitem{korzh2020demonstration}
B.~Korzh, Q.-Y. Zhao, J.~P. Allmaras, S.~Frasca, T.~M. Autry, E.~A. Bersin,
  A.~D. Beyer, R.~M. Briggs, B.~Bumble, M.~Colangelo \emph{et~al.},
  \enquote{Demonstration of sub-3 ps temporal resolution with a superconducting
  nanowire single-photon detector,} {Nature Photonics}
  \textbf{14}, 250--255 (2020).

\bibitem{wiersig2009direct}
J.~Wiersig, C.~Gies, F.~Jahnke, M.~A{\ss}mann, T.~Berstermann, M.~Bayer,
  C.~Kistner, S.~Reitzenstein, C.~Schneider, S.~H{\"o}fling \emph{et~al.},
  \enquote{Direct observation of correlations between individual photon
  emission events of a microcavity laser,} {Nature}
  \textbf{460}, 245--249 (2009).

\bibitem{assmann2010ultrafast}
M.~A{\ss}mann, F.~Veit, M.~Bayer, C.~Gies, F.~Jahnke, S.~Reitzenstein,
  S.~H{\"o}fling, L.~Worschech, and A.~Forchel, \enquote{Ultrafast tracking of
  second-order photon correlations in the emission of quantum-dot
  microresonator lasers,} {Physical Review B}
  \textbf{81}, 165314 (2010).

\bibitem{assmann2010measuring}
M.~A{\ss}mann, F.~Veit, J.-S. Tempel, T.~Berstermann, H.~Stolz, M.~van~der
  Poel, J.~M. Hvam, and M.~Bayer, \enquote{Measuring the dynamics of
  second-order photon correlation functions inside a pulse with picosecond time
  resolution,} {Optics Express} \textbf{18},
  20229--20241 (2010).

\bibitem{venkatesh2016ultrafast}
Y.~Venkatesh, M.~Venkatesan, B.~Ramakrishna, and P.~R. Bangal,
  \enquote{Ultrafast time-resolved emission and absorption spectra of
  meso-pyridyl porphyrins upon soret band excitation studied by fluorescence
  up-conversion and transient absorption spectroscopy,}
 {The Journal of Physical Chemistry B} \textbf{120},
  9410--9421 (2016).

\bibitem{kandori1994direct}
H.~Kandori, H.~Sasabe, and M.~Mimuro, \enquote{Direct determination of a
  lifetime of the s$_2$ state of $\beta$-carotene by femtosecond time-resolved
  fluorescence spectroscopy,} {Journal of the American
  Chemical Society} \textbf{116}, 2671--2672 (1994).

\bibitem{chen2012metallodielectric}
X.-W. Chen, M.~Agio, and V.~Sandoghdar, \enquote{Metallodielectric hybrid
  antennas for ultrastrong enhancement of spontaneous emission,}
 {Physical Review Letters} \textbf{108}, 233001 (2012).

\bibitem{hoang2016ultrafast}
T.~B. Hoang, G.~M. Akselrod, and M.~H. Mikkelsen, \enquote{Ultrafast
  room-temperature single photon emission from quantum dots coupled to
  plasmonic nanocavities,} {Nano Letters} \textbf{16},
  270--275 (2016).

\bibitem{flatae2019plasmon}
A.~M. Flatae, F.~Tantussi, G.~C. Messina, F.~De~Angelis, and M.~Agio,
  \enquote{Plasmon-assisted suppression of surface trap states and enhanced
  band-edge emission in a bare cdte quantum dot,} {The
  Journal of Physical Chemistry Letters} \textbf{10}, 2874--2878 (2019).

\bibitem{bogdanov2020ultrafast}
S.~I. Bogdanov, O.~A. Makarova, X.~Xu, Z.~O. Martin, A.~S. Lagutchev,
  M.~Olinde, D.~Shah, S.~N. Chowdhury, A.~R. Gabidullin, I.~A. Ryzhikov
  \emph{et~al.}, \enquote{Ultrafast quantum photonics enabled by coupling
  plasmonic nanocavities to strongly radiative antennas,}
  {Optica} \textbf{7}, 463--469 (2020).

\bibitem{zhang2006femtosecond}
L.~Zhang, Y.-T. Kao, W.~Qiu, L.~Wang, and D.~Zhong, \enquote{Femtosecond
  studies of tryptophan fluorescence dynamics in proteins: Local solvation and
  electronic quenching,} {The Journal of Physical
  Chemistry B} \textbf{110}, 18097--18103 (2006).

\bibitem{jimenez1994femtosecond}
R.~Jimenez, G.~R. Fleming, P.~Kumar, and M.~Maroncelli, \enquote{Femtosecond
  solvation dynamics of water,} {Nature} \textbf{369},
  471--473 (1994).

\bibitem{barth2007infrared}
A.~Barth, \enquote{Infrared spectroscopy of proteins,}
  {Biochimica et Biophysica Acta (BBA)-Bioenergetics}
  \textbf{1767}, 1073--1101 (2007).

\bibitem{mao2015multi}
P.~Mao, Z.~Wang, W.~Dang, and Y.~Weng, \enquote{Multi-channel lock-in amplifier
  assisted femtosecond time-resolved fluorescence non-collinear optical
  parametric amplification spectroscopy with efficient rejection of
  superfluorescence background,} {Review of Scientific
  Instruments} \textbf{86}, 123113 (2015).

\bibitem{sajadi2013femtosecond}
M.~Sajadi, M.~Quick, and N.~Ernsting, \enquote{Femtosecond broadband
  fluorescence spectroscopy by down-and up-conversion in $\beta$-barium borate
  crystals,} {Applied Physics Letters} \textbf{103},
  173514 (2013).

\bibitem{Kim:08}
C.~H. Kim and T.~Joo, \enquote{Ultrafast time-resolved fluorescence by two
  photon absorption excitation,} {Opt. Express}
  \textbf{16}, 20742--20747 (2008).

\bibitem{tan2008control}
W.~Tan, H.~Liu, J.~Si, and X.~Hou, \enquote{Control of the gated spectra with
  narrow bandwidth from a supercontinuum using ultrafast optical kerr gate of
  bismuth glass,} {Applied Physics Letters} \textbf{93},
  051109 (2008).

\bibitem{schmidt2003broadband}
B.~Schmidt, S.~Laimgruber, W.~Zinth, and P.~Gilch, \enquote{A broadband kerr
  shutter for femtosecond fluorescence spectroscopy,}
  {Applied Physics B} \textbf{76}, 809--814 (2003).

\bibitem{kinoshita2000efficient}
S.~Kinoshita, H.~Ozawa, Y.~Kanematsu, I.~Tanaka, N.~Sugimoto, and S.~Fujiwara,
  \enquote{Efficient optical kerr shutter for femtosecond time-resolved
  luminescence spectroscopy,} {Review of Scientific
  Instruments} \textbf{71}, 3317--3322 (2000).

\bibitem{saleh2019fundamentals}
B.~E. Saleh and M.~C. Teich, \emph{Fundamentals of Photonics} (John Wiley \&
  Sons, 2019).

\bibitem{doi:10.1366/0003702053085007}
S.~Arzhantsev and M.~Maroncelli, \enquote{Design and characterization of a
  femtosecond fluorescence spectrometer based on optical kerr gating,}
  {Applied Spectroscopy} \textbf{59}, 206--220 (2005).
  PMID: 15720762.

\bibitem{lightcon}
\enquote{Optics toolbox - interactive calculators for scientists and
  engineers,} \url{http://toolbox.lightcon.com/tools/dispersionparameters/}.
  Accessed: 2020-10-13.

\bibitem{karras2014nonlinear}
C.~Karras, D.~Litzkendorf, S.~Grimm, K.~Schuster, W.~Paa, and H.~Stafast,
  \enquote{Nonlinear refractive index study on
  {SiO$_2$-Al$_2$O$_3$-La$_2$O$_3$} glasses,} {Optical
  Materials Express} \textbf{4}, 2066--2077 (2014).

\end{thebibliography}
\end{document}